\documentclass[sigconf]{acmart}
\usepackage{enumitem}
\setlist{listparindent=1.25em}
\usepackage[figuresright]{rotating}
\usepackage{tabularx}
\usepackage{balance}
\usepackage{tikz}
\usepackage{pgfplots,pgfplotstable}
\usepgfplotslibrary{statistics}
\usepackage{multirow}
\usepackage[normalem]{ulem}
\useunder{\uline}{\ul}{}
\usetikzlibrary{shapes.geometric, arrows, positioning, patterns}
\newcommand\identifying[1]{#1}

\usepackage{newfloat}
\DeclareFloatingEnvironment[
    fileext=loa,
    listname={List of Appendices},
    name=Appendix,
    placement={!ht},
    within=none,
]{floatappendix}

\AtBeginDocument{%
  \providecommand\BibTeX{{%
    \normalfont B\kern-0.5em{\scshape i\kern-0.25em b}\kern-0.8em\TeX}}}

\acmConference[arXiv Pre-print]{arXiv Pre-print}{}{}
\acmBooktitle{arXiv Pre-print}
\renewcommand\footnotetextcopyrightpermission[1]{}
\makeatletter
\renewcommand\@formatdoi[1]{\ignorespaces}
\makeatother

\begin{document}
\fancyfoot{}
\fancyhead{}
\title[Selective Marketing to Broaden Participation in CS Education]{Assessing the Effectiveness of Selective Marketing to Broaden Participation in CS Education}
\pagestyle{plain}
\author{Aditya Shah}
\email{adshah@uw.edu}
\orcid{0009-0002-4068-135X}
\affiliation{
 \institution{University of Washington}
 \city{Seattle}
 \state{Washington}
 \country{USA}
}

\author{Tyler Menezes}
\email{tylermenezes@codeday.org}
\orcid{0000-0002-7975-2533}
\affiliation{
 \institution{CodeDay}
 \city{Seattle}
 \state{Washington}
 \country{USA}
}

\begin{abstract}
Many studies have aimed to broaden participation in computing (BPC) through extracurricular educational initiatives. When these initiatives are structured as open-enrollment extracurricular programs, their success often depends on their marketing approach. However, there is little in the computing education research literature about how to conduct effective marketing for these initiatives. We describe the changes made to the marketing strategy of one such program, an educational hackathon for middle school and high school students in the Pacific Northwest. These included reducing promotion to affluent families, using targeted school-based communication, and emphasizing cost supports during initial promotion. We then compare attendance and self-reported demographics before and after the intervention. Results indicate a higher proportion of students from marginalized and low-income communities and no reduction in overall attendance.
\end{abstract}

\maketitle
\section{Introduction}

Extracurricular, open-enrollment computing education events are often promoted through channels that disproportionately reach students from affluent backgrounds and well-resourced communities, leading to participant pools that underrepresent marginalized groups. This inequality limits access to the social, educational, and career benefits these events provide and risks reinforcing existing gaps in technology employment. \identifying{CodeDay} is one such programming event that has been studied for its potential to improve diversity. \identifying{\cite{ganHowGenderEthnicity2022}} 
The objective of this research is to test whether an improved promotion strategy can help broaden participation in extracurricular, in-person CS education programs and provide practical guidance for inclusive recruitment in computer science education contexts. Although the influence of marketing may seem intuitive, BPC research rarely isolates it from program design, making it important to measure its independent impact. We test the hypothesis that \textbf{(H1)} limited promotion in affluent neighborhoods, leveraged teacher and community partner channels, and highlighted cost supports will improve the proportion of underrepresented students participating in the program.

\section{Prior Work}

K-12 CS education programs use a range of marketing and recruitment strategies. Among extracurricular and multi-school programs, studies have discussed active recruitment practices such as letter-writing campaigns and recruiting from introductory CS courses, \cite{karlinBuildingGenderinclusiveSecondary2023} alongside sustained partnerships between schools, districts, universities, and industry. \cite{goodeIncreasingDiversityK122008} Several programs have addressed diversity by offering professional development opportunities for strategically selected schools, student recruitment efforts, volunteer training, and curriculum design.\cite{ibeReflectionsDiversityEquity2018,goodeIfYouBuild2007} Intra-school studies have found that teacher support from counselors and administrators is essential, with successful programs requiring holistic, schoolwide stakeholder involvement rather than relying solely on CS teachers.\cite{karlinBuildingGenderinclusiveSecondary2023,gretterEquitableLearningEnvironments2019} Universal Design for Learning principles have also been implemented to create accessible programming environments.\cite{hansenDifferentiatingDiversityUsing2016}

These studies all describe thoughtful recruitment practices. However, there is limited literature examining the effect size of particular marketing practices. It is difficult to determine how much of the success of BPC programs in recruiting diverse participants is related to appealing program design versus marketing tactics.

\section{Methods}

Prior to the intervention, program marketing comprised (1) technology companies sharing event information with their employees and their families; and (2) promotional flyers provided for distribution to all local high schools. Events included a nominal registration fee of \$5-\$10; while the fee was intended to offset costs, scholarships were offered without restriction to any student who requested one, with discounts being provided proactively to technology companies and schools.

To study the effect size of the marketing strategy, we revised these marketing tactics to reduce marketing to affluent areas, amplify communication channels that reach lower-income communities, and provide details about cost supports upfront in marketing materials: 

1. \textbf{Reducing promotion to affluent families}. To avoid skewed attendance from affluent neighborhoods, the event was not advertised through channels disproportionately accessed by higher-income families. We ceased flyer distribution to affluent schools and promotion through corporate partners. Additionally, we did not announce events on the \identifying{CodeDay} social media accounts, because these followers were more affluent and familiar with the event.

2. \textbf{Targeted school-based communication}. A comprehensive list of Title I high schools across the three nearest counties was developed, and outreach messages were distributed to teachers across a range of subject areas (including arts, humanities, and college/career readiness) rather than exclusively computer science departments. This approach was designed to reach students who may not already have identified with computer science as an academic pathway. Promotional posts were also shared via school-affiliated social media accounts and e-flyer services when available.

3. \textbf{Emphasis on cost supports}. Outreach materials explicitly highlighted cost-reducing supports and attendance benefits, such as free transportation options and a free laptop. Messaging was framed to lower perceived financial and logistical barriers to attending the events.

\section{Results}

Data were collected via registration forms at two events prior to the implementation of the revised promotional strategy, in summer and fall of 2023, and two events following implementation, in winter and spring of 2024. ($N = 285; n_{pre} = 121, n_{post} = 164$). We analyzed participant demographics according to gender, race/ethnicity, and socioeconomic status. Each variable was dichotomized to reflect participation by underrepresented groups, aligning with our focus on barriers broadly rather than subgroup outcomes:

\textbf{Gender identity.} Coded as \textit{Male} (he/him) or \textit{NonMale} (others).

\textbf{Race/ethnicity.} Coded as \textit{Underrepresented} (non-White) or \textit{Not\-Underrepresented} (White), reflecting the demographic composition of software engineers in the studied region and time period.

\textbf{Socioeconomic status.} Average household income by residential postal code, as retrieved from American Community Survey 5-year estimates, was used as a proxy due to challenges in collecting individual income data from minors. Participants were coded as \textit{LowIncome} (median household income $< 150\%$ of the Federal poverty line) or \textit{NotLowIncome} ($\ge 150\%$ of the Federal poverty line).

We conducted Pearson's chi-squared tests with Yates' continuity correction to examine associations between promotional strategy and each demographic variable. To address the issue of multiple comparisons, we applied the Holm–Bonferroni correction method to adjust p-values. For associations reaching statistical significance, we calculated Cramer's V as a measure of effect size. 

We found an association between the improved promotion strategy and both ethnicity ($\chi^2=79.7, p<0.0001$) and income level ($\chi^2=65.6, p<0.0001$) but not gender ($\chi^2=1.17, p=0.28$). In the improved promotion strategy group, there were more students than expected under the hypothesis of independence who are from underrepresented ethnic ($stdres=9.04$) and low-income ($stdres=8.22$) backgrounds. According to our results, there is a large association between the improved promotion strategy and ethnicity (Cramer's V = $0.53$) and a medium association with income level (V = $0.49$).

\section{Discussion}

Our findings demonstrate that targeted promotional strategies can substantially improve participation from students of underrepresented racial/ethnic backgrounds and low-income communities in extracurricular K-12 computing education programs. These results support H1 in emphasizing the importance of active, intentional recruitment strategies. Teachers and other community partners likely act as cultural brokers who communicate the value of CS programs to families who might otherwise see them as "not for them." Additionally, highlighting cost supports upfront may reduce uncertainty about perceived hidden costs. The focus on inclusion and barrier removal is critical in addressing the factors that contribute to attrition from STEM fields, such as computer science, where underrepresented students are more likely to leave the field due to stereotype threat.\cite{beasleyWhyTheyLeave2012} The lack of significant gender differences may indicate that gender-based barriers to participation operate through different mechanisms from those addressed by our intervention and require further investigation.

\section{Threats to Validity and Future Work}

The pre–post design without a control group leaves open the possibility of temporal or concurrent influences. Additionally, the findings from a single geographic location may have limited generalizability to programming events in regions with different demographic compositions or where the baseline levels of representation of underrepresented groups in computing differ. Using ZIP code-level income proxies and dichotomized demographic categories may oversimplify participants’ lived experiences.

\bibliographystyle{format/ACM-Reference-Format}
\balance
\bibliography{arxiv}

\end{document}